# Impact of Local Stiffness on Entropy Driven Microscopic Dynamics of Polythiophene

Sudipta Gupta, [1,*] Sourav Chatterjee, [1] Piotr Zolnierczuk, [2] Evgueni E. Nesterov, [1,3*] and Gerald J. Schneider[1,4*]

[1]*Department of Chemistry, Louisiana State University, Baton Rouge, LA 70803, USA*

[2]*Jülich Centre for Neutron science (JCNS) and outstation at SNS, Oak Ridge, POB 2008, 1 Bethel Valley Road, TN 37831, USA*

[3]*Department of Chemistry and Biochemistry, Northern Illinois University, DeKalb, IL 60115, USA*

[4]*Department of Physics and Astronomy, Louisiana State University, Baton Rouge, LA 70803, USA*

## Abstract

We exploited the high temporal and spatial resolution of neutron spin echo spectroscopy to investigate the large-scale dynamics of semiflexible conjugated polymer chains in solutions. We obtained the first experimental demonstration of earlier predicted single chain glassy state. We used a generalized approach of the well-established Zimm model of flexible polymers to describe the relaxation mode spectra of locally stiff polythiophene chains. The Zimm mode analysis confirms the existence of beads with a finite length that corresponds to a reduced number of segmental modes in semiflexible chains. Irrespective of the temperature and the molecular weight of the conjugated polymer, we witness a universal behavior of the local chain stiffness and invariability of the beads length. Our experimental findings indicate possibly minor role of the change in π-electron conjugation length (and therefore conjugated backbone planar to non-planar conformational transition) in the observed thermochromic behavior of polythiophene but instead point on the major role of chain dynamics in this phenomenon.

## Introduction

Since the pioneering work of Rouse, Zimm, and De Gennes[1-3] virtually all of our current understanding of the large-scale dynamics of polymers is based on flexible polymers. However, most biopolymers, like DNA, cellulose,[4,5] and advanced electronic polymers, such as conjugated polymers,[6] are locally quite stiff. On length-scales shorter than the Kuhn length, the relaxation mode spectra of semiflexible and flexible chains are considerably different. This has a substantial impact on macroscopic properties, such as the response to mechanical, electrical, and optical stimulation as well as on proper functioning of the living cell. A unified approach that self-consistently describes the large-scale chain dynamics for both flexible and semiflexible chains is of paramount importance to understand the microscopic origin of their macroscopic properties. We exploit the high temporal and spatial resolution of neutron spin echo (NSE) spectroscopy to identify a unique set of physical parameters that can quantify the impact of the chain stiffness on the large-scale polymer dynamics.

In case of flexible polymers at the length-scale larger than monomer size, the chemical details become unimportant and the statistical nature of the chain gives rise to the conformational entropic forces. The large-scale dynamics of the polymer is driven by the balance between thermal fluctuations and the entropy-driven restoring forces. The force constant is inherently connected to the flexibility of the chain. In semiflexible polymers, a finite damping slows down the relaxation towards achieving thermodynamic equilibrium. Such forces are responsible for the rubber elasticity.[7-11] The finite rigidity of the chains is responsible for nonaffine rubber elasticity in cross-linked polymer networks[12] and shear-induced banding in polymeric liquids.[13]

Depending on the solvent quality and temperature, long flexible polymers in dilute solution can assume a swollen coil, a Gaussian chain, or collapse to a globule conformation. On the contrary, semiflexible polymers can exhibit random coil to rod-like conformations. In dilute solutions, the long-range hydrodynamic interaction imparted by the solvent molecules reduces frictional resistance and exhibits faster chain dynamics. The inherent stiffness reduces the segmental mobility and can lead to a partially frozen or glassy state. De Gennes suggested that this affects the mode spectrum of single chains.[14] Recently, NSE spectroscopy experiments on semiflexible polymer chains suggested the existence of such single chain glassy (SCG) states.[15,16] The organic polymeric semiconductors represent a unique class, in which the delocalized π-electrons essentially govern the chain stiffness and the macroscopic optoelectronic properties.[17-21] At the same time, chemical defects[22] and dynamic heterogeneity[23] in conjugated polymers cause a reduction in conductivity and change in other electronic properties.[24] Recent small angle neutron scattering (SANS) studies on poly(3-alkylthiophene)s suggest a close relationship between the polymer conformation and its optoelectronic characteristics, such as those related to thermochromism phenomenon.[25,26] Nevertheless, understanding the role of dynamics of conjugated polymer chain on its electronic and optical characteristics remains elusive.

In this work, we report a comprehensive study of the temperature-dependent single chain conformation and dynamics of regioregular poly(3-hexylthiophene) (P3HT) in deuterated 1,2-dichlorobenzene (DCB-$D_4$) solvent using SANS and NSE. P3HT was chosen as a well-studied representative of conjugated polymers that has proven to be an archetypal material for electronic and optoelectronic applications, and is often used to understand properties of conjugated polymers in general.[27-29] The materials and the systems are described in Section 2, together with the theoretical and experimental techniques applied to the analysis of the results. The latter are presented in Section 3 and the obtained dynamic data are interpreted in terms of the static structural properties derived by complimentary SANS experiments. An extensive discussion is presented in Section 4, whereas in Section 5 we summarize and draw our conclusions.

**Results**

We carried out experiments on two P3HT samples of different molecular weights ($M_n$ 63.1 kg/mol and 89.7 kg/mol), which had close to 100 % regioregularity (Table 1). Experiments were conducted at elevated temperatures and a low concentration ($\phi$ = 0.75 %) to reduce inter-chain aggregation.[30]

**Polymer synthesis.** P3HT samples with number-average molecular weight $M_n$ 63.1 kg/mol and 89.7 kg/mol were prepared using controlled Kumada catalyst-transfer polymerization following the procedure described in reference.[31] The preparation was carried out using an external catalytic initiator prepared via the reaction of 5-bromo-2,2′-bithiophene and bis[1,3-bis(diphenylphosphino)propane]nickel(0), and the polymer samples with different molecular weights were obtained through the variation of the ratio between the external catalytic initiator and 5-bromo-4-hexyl-2-thienylmagnesium chloride monomer. The polymers were additionally purified using Soxhlet extraction, with successive extraction with methanol, hexane, and $CHCl_3$. Determination of $M_n$ and polydispersity index (PDI) was carried out with GPC (using Agilent 1100 chromatograph equipped with two PLgel 5 μm MIXED-C and one PLgel 5 μm 1000 Å columns connected in series, using THF as a mobile phase) calibrated against polystyrene standards.

Regioregularity of the P3HT samples was determined using $^1$H NMR spectroscopy (Figure 1) as described in reference.[31]

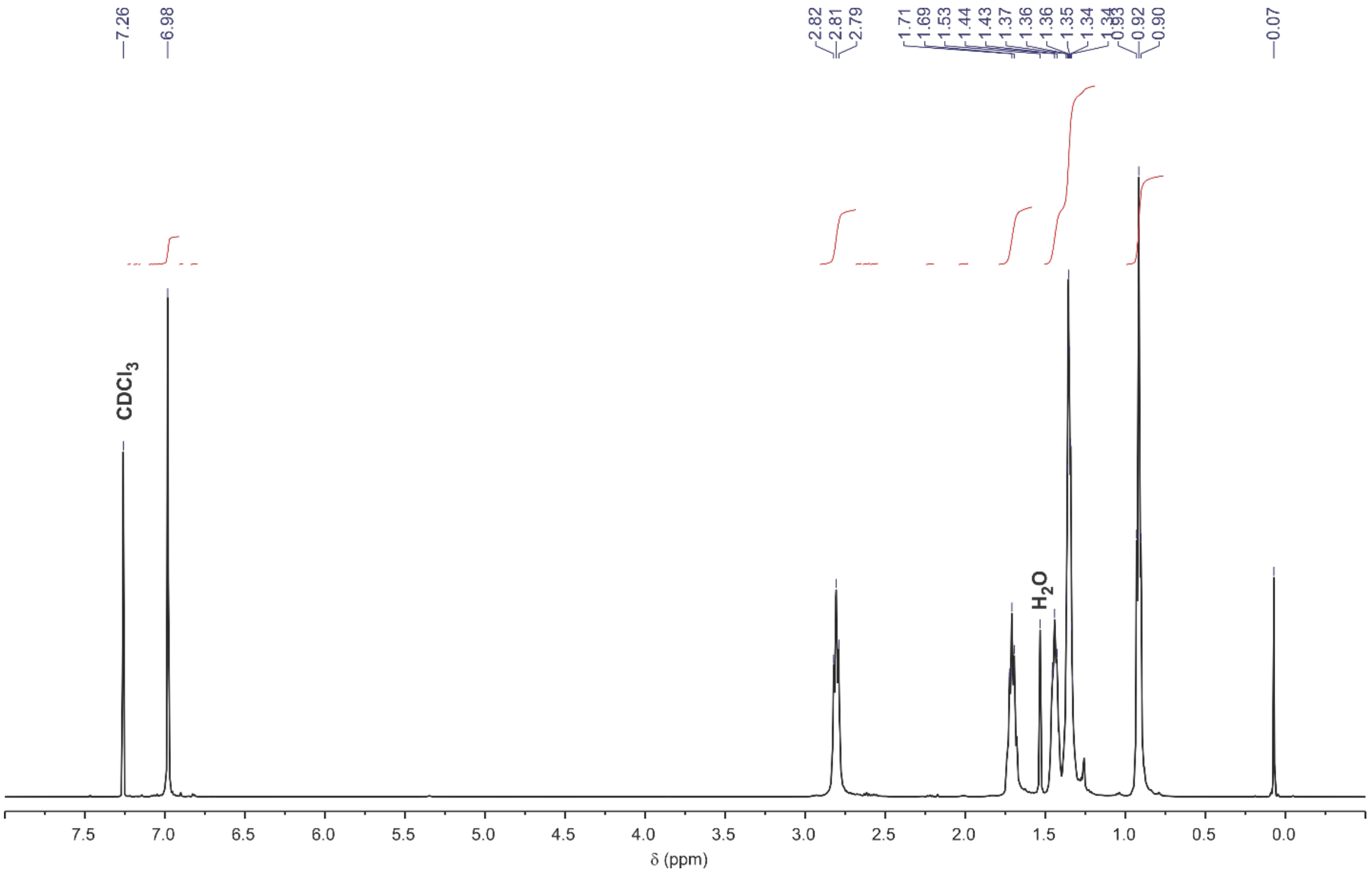

*Figure 1. Representative $^1$H NMR spectrum of P3HT (63.1 kg/mol) in $CDCl_3$.*

In order to minimize aggregates or chains folding upon themselves owning to their strong intramolecular π-π stacking interactions,[30] solutions of P3HT in DCB-$D_4$ were prepared via heating and stirring P3HT samples with the solvent at 70 °C overnight. In every neutron scattering experiment, the samples were equilibrated at constant temperature for 30 mins in a tumbler before measurements were performed.

**Theoretical description.** Our theoretical approach is based on detailed analysis of the spectrum of relaxation modes to account for the entropic forces and hydrodynamic interactions of polymer in solutions. NSE spectroscopy measures the normalized dynamic structure factor $S(Q,t)/S(Q)$ as a function of Fourier time, $t$ at a given momentum transfer, $Q$. At the intermediate length scale, the center of mass diffusion and segmental relaxation of a polymer melt is well described by the Rouse model. The molecular motion originates from the balance between entropic and frictional forces caused by the surrounding heat bath and is best described by its spectrum of relaxation modes.[32] For polymers in solution, the hydrodynamic interactions become important, and the dynamic structure factor can be formulated within the framework of the Zimm model:[1]

$$S_{chain}(Q,t) = \frac{1}{N}\exp[-Q^2 D_Z t]\sum_{n,m}\exp\left\{-\frac{1}{6}|n-m|^{2\nu}Q^2\ell^2\right\}\times$$
$$\exp\left\{-\frac{2}{3}\frac{R_{ee}^2 Q^2}{\pi^2}\sum_p \frac{1}{p^{2\nu+1}}\cos\left(\frac{p\pi m}{N}\right)\cos\left(\frac{p\pi n}{N}\right)\left(1-\exp\left(-\frac{tp^{3\nu}}{\tau_z}\right)\right)\right\} \tag{1}$$

Here $n$, $m$ are the polymer segment numbers where the summation runs over the total number of monomer segments, $N$. The statistical segment length is given by $\ell$ and is obtained from $R_{ee}^2 = \ell^2 N^{2\nu}$. The first part in equation 1 describes the Zimm center of mass diffusion with a diffusion coefficient $D_Z = \frac{\alpha_D k_B T}{(\eta_s R_{ee})}$, here, $\eta_s$ is the solvent viscosity. The constant pre-factor, $\alpha_D$ = 0.196 (Θ solvent) and $\alpha_D$ = 0.203 (good solvent).[33] $S_{chain}(Q)$ represents the static structure factor of the chain. The third term represents the more local dynamics, including rotational diffusion ($p$ = 1). It is represented by a sum over relaxation modes of the polymer chain with mode number $p$, and characteristic time $\tau_p = \tau_Z p^{-3\nu}$. The corresponding Zimm segmental relaxation time is given by $\tau_Z = 0.325\eta_s R_{ee}^3/(k_B T)$,[33] with $\eta_s$ being the solvent viscosity at a thermal energy $k_B T$, where $k_B$ is the Boltzmann constant.

**Chain conformation.** We determined the unperturbed chain dimensions by SANS experiments. The scattering data, intensity vs. momentum transfer, $Q$, can be found in the SI. In these data, we see a typical form factor of an aggregated polymer. The Flory exponent and the radius of gyration, $R_g$, can be conveniently extracted from the Kratky plot, as illustrated in Figure 2. For the sake of clarity we omitted the intensity values at low $Q$.[34] Figure 2 displays Kratky plots for the two polymers with two different molecular weights at two temperatures in DCB-$D_4$ solutions. The plateau at intermediate momentum transfer, $Q$, is a signature of a Gaussian coil and indicates a scaling relationship $I \sim Q^{-1/\nu} = Q^{-2}$.[34] For a Θ solvent, the Flory exponent, $\nu$ = ½, and for a good solvent, $\nu$ = 0.588. Thus, SANS data verify random coils in a Θ solvent. A fit with the Debye function: $\frac{2}{(u)^2}[u - 1 + \exp(-u)]$, with $u = Q^2 {R_g}^2$, shows an increasing radius of gyration, $R_g$, or chain end-to-end distance, $R_{ee} = \sqrt{6}R_g$, with increasing molecular weight and a slight variation with temperature (Table 1). It shows only a minor change with both molecular weight and temperature. Slight deviations of the fit at high $Q$ are due to the incoherent background scattering that increases the noise level, but do not change our results on $R_g$. Since, ${R_g}^2 \propto M_n$, which is valid within experimental accuracy as shown in Table 1. This result is in favor of our assumption of Gaussian statistics of the chain with stiff segments. We observed an increase of $R_g$ with temperature by 10 to 15%.

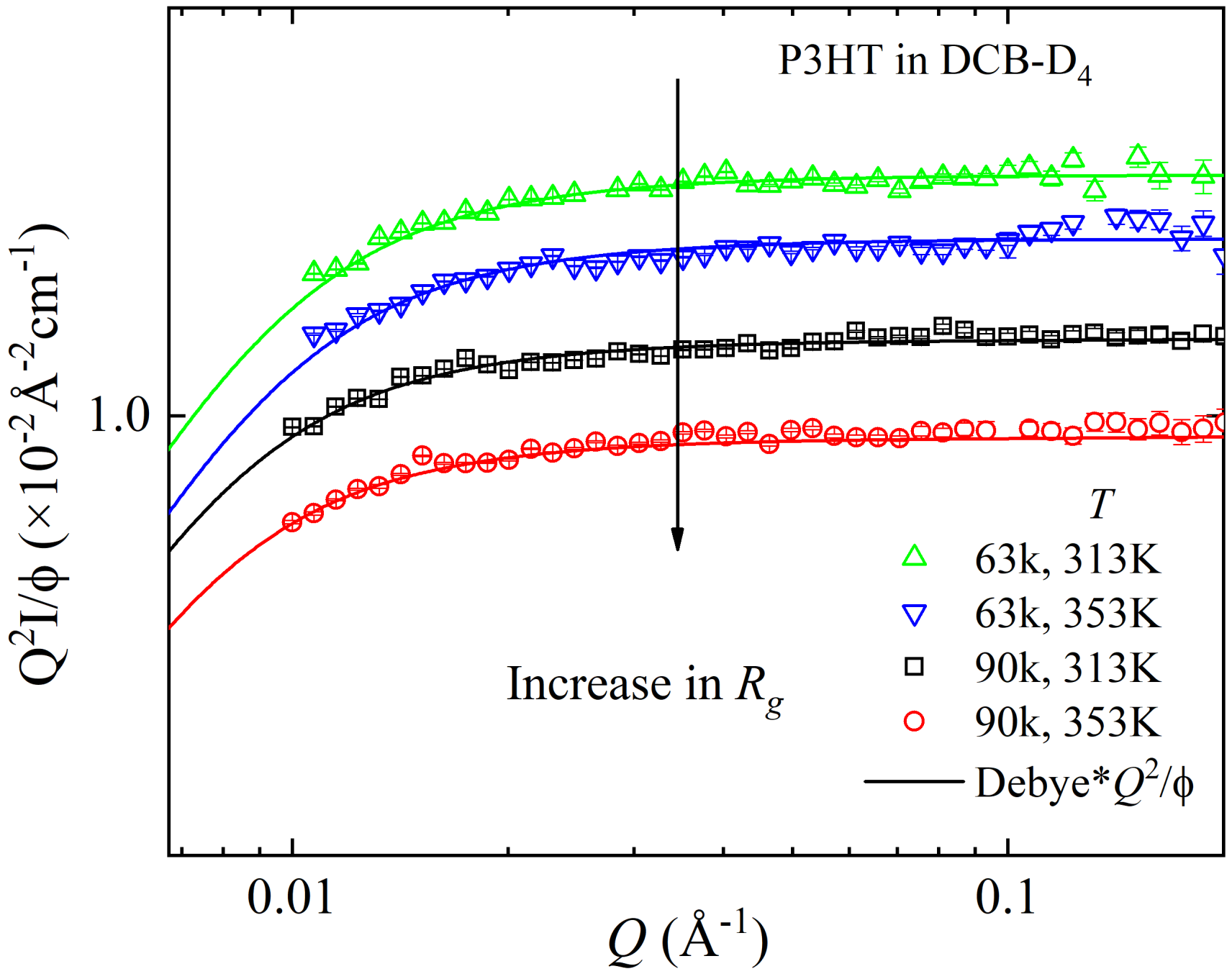

*Figure 2. Kratky plots for P3HT of two different molecular weights and temperatures in DCB-$D_4$, determined from SANS. The data are normalized by their volume fraction, $\phi$. The lines represent the fit with Debye model. The original intensity data can be found in the SI.*

*Table 1. Sample labels, molecular weight $M_n$, polydispersity $M_w/M_n$ (determined by $^1$H NMR and GPC), radius of gyration $R_g$ (determined from SANS). The chain end-to-end distance, $R_{ee}$, Zimm diffusion, $D_Z$, and the Zimm time, $\tau_z$, are calculated. The Zimm modes, p, the estimated number of modes, $p_{min}$, above which the NSE relaxation spectra is independent of p, the stiffness parameter, $\alpha$, and the dynamic rigid length, $R_{rigid}$, as obtained from NSE experiments.*

| P3HT Label | $M_n$ (kg/mol) | PDI = $M_w/M_n$ | $T$ (K) | $R_g$ (nm) | $R_{ee}$ (nm) | $D_z$ × 10$^{-2}$ (nm$^2$ ns$^{-1}$) | $\tau_z$ (ns) | $p$ | $p_{min}$ | α | $R_{rigid}$ (nm) |
|---|---|---|---|---|---|---|---|---|---|---|---|
| 63k | 63.1 | 1.51 | 313 | 15±1 | 37±2 | 2.36 | 4125 | 15±1 | 60 | 0.030±0.007 | 4.77±0.09 |
| 63k | 63.1 | 1.51 | 353 | 17±1 | 41±3 | 3.43 | 3195 | 18±2 | 75 | 0.022±0.003 | 4.72±0.05 |
| 90k | 89.7 | 1.54 | 313 | 19±1 | 48±2 | 1.67 | 8884 | 27±2 | 100 | 0.013±0.006 | 4.77±0.07 |
| 90k | 89.7 | 1.54 | 353 | 21±1 | 50±3 | 2.70 | 5998 | 27±3 | 120 | 0.008±0.001 | 4.60±0.13 |

**Chain dynamics.** Figure 3 illustrates $S(Q,t)/S(Q)$ obtained by NSE experiments over a $Q$-range from 0.062 to 0.124 Å$^{-1}$, for P3HT of two different molecular weights, at 313K and 353K.

First, we assumed a rigid polymer model and calculated $D_Z$ and $\tau_Z$ from the solvent viscosity $\eta_s$, and the chain end-to-end distance $R_{ee}$ as obtained independently from SANS (Table 1). As can be seen, this model does not suffice to describe the measured data (cf. SI). The much faster decay of our experimental data indicates a substantial contribution of another relaxation mechanism.

In a next step, we considered P3HT in solution as a rigid worm like chain as proposed by McCulloch *et al.*,[25] which requires to add the rotational diffusion ($p = 1$). The comparison with the experimental data shows that this is still not sufficient (cf. SI). Hence, we improve the model by considering a polymer coil with mobile segments, that requires to include the segmental relaxation ($p > 1$). We obtained an accurate description of $S(Q,t)$ by adding only a finite number of modes, $p = 2, \ldots, P$, cf. Figures 3 (a) and (b). The number of modes needed is surprisingly low, with $P$ ranging from 15 to 27. Within the experimental accuracy, the $p$ is temperature independent but changes with $M_n$. This observation is expected because the number of modes is proportional to the number of repeating units in the polymer chain.[35] The corresponding analysis protocol for different modes is presented in the SI.

The parameter, $P$, represents the number of modes necessary to describe the experimental dynamic structure factor $S(Q,t)/S(Q)$ at different temperatures and molecular weights simultaneously for all $Q$'s. We would like to emphasize that this theoretical description of the experimental NSE data involves no free parameter, except $P$.

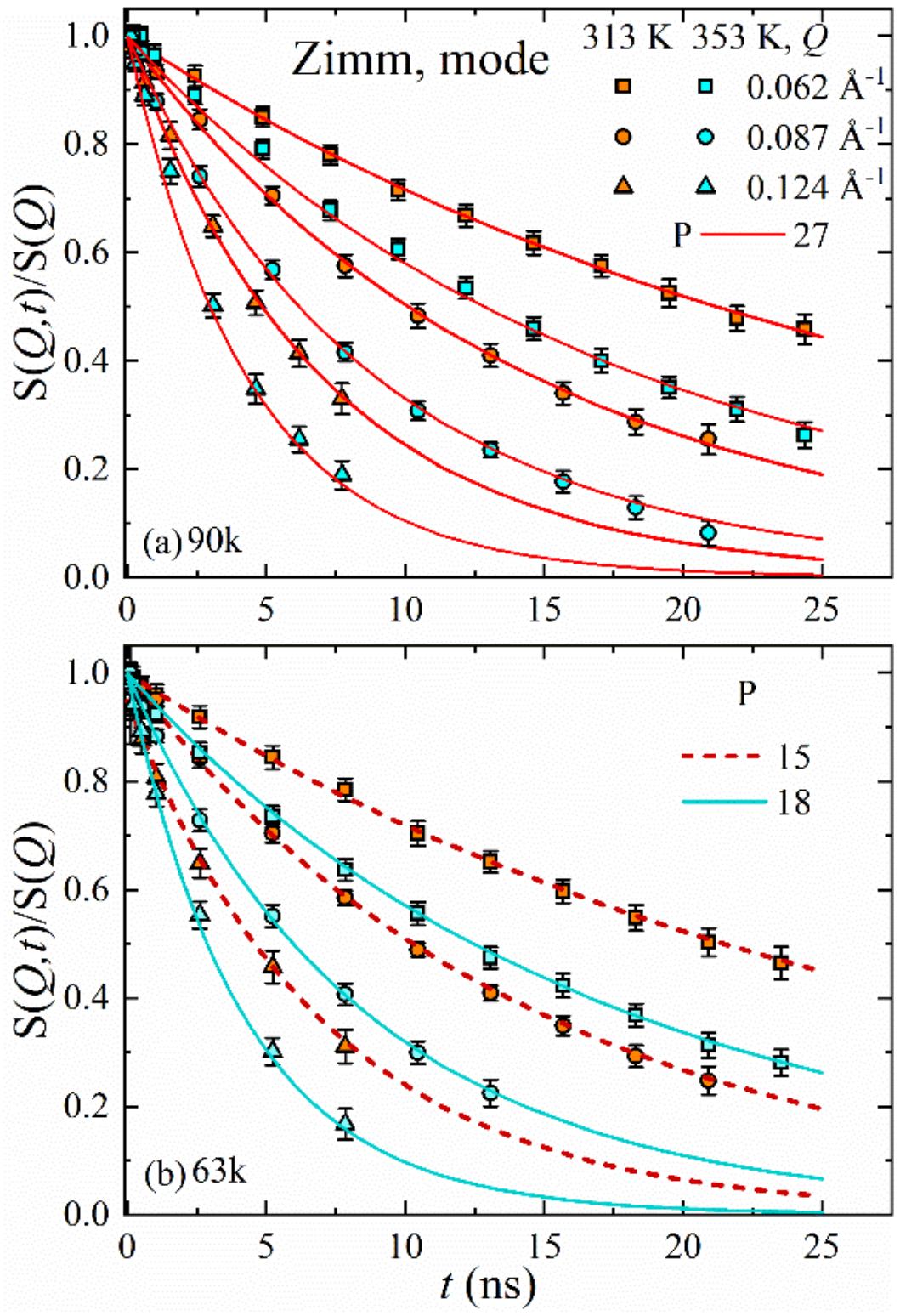

*Figure 3. Normalized dynamic structure factor, $S(Q,t)/S(Q)$, as a function of Fourier time, t, for P3HT samples with different $M_n$ = 90 (top) and 63 kg/mol (bottom) and T = 313K (red) and 353K (cyan), respectively. The different momentum transfer, Q, is given by, □ = 0.062 $Å^{-1}$, ○ = 0.087 $Å^{-1}$ and △ = 0.124 $Å^{-1}$. The solid and dashed lines in (a) and (b) represent the summation over finite number of Zimm modes, P.*

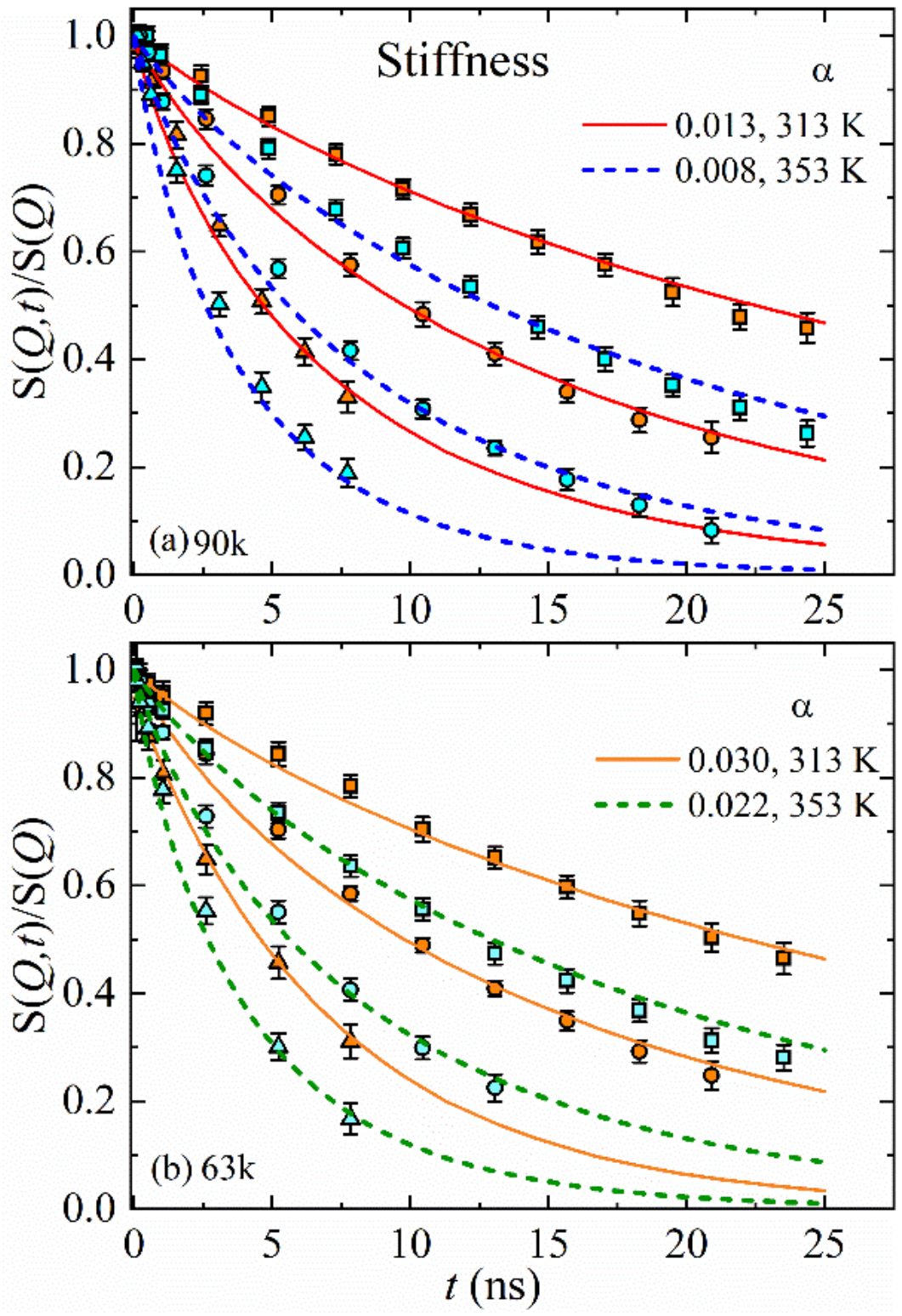

*Figure 4. Normalized dynamic structure factor, $S(Q,t)/S(Q)$, as a function of Fourier time, t, for P3HT samples with different molecular weight (a) $M_n$ = 90, and (b) 63kg/mol. Each figure includes two temperatures, T = 313K (solid) and 353K (dash). The different momentum transfer, Q, is given by, □ = 0.062 $Å^{-1}$, ○ = 0.087 $Å^{-1}$ and △ = 0.124 $Å^{-1}$. The solid and dashed lines in (a) and (b) represent the best fit summing over a large number of modes, $p_{min}$, and a finite dynamical stiffness, $\alpha$.*

**Discussions**

Limiting the analysis to finite number of modes *P* ignores a substantial part of the mode spectrum and seems to be unjustified. On the other hand, the increased stiffness caused by the delocalized π-electron system in P3HT introduces a finite correlation length (dynamic equivalent to the static Kuhn segment), which can be taken into account by adding a fourth-order term, $p^2 + \alpha p^4$, to the entropic spring constant ($k = 3k_BT/\ell^2 \propto p^{-2}$), with the dynamic stiffness parameter $\alpha$.[15,36,37] The modified Zimm scattering model is obtained by replacing the mode dependence, $p^{3\nu}$, of $\tau_z$ (in equation 1) by $p^{3\nu} + \alpha p^{4-\nu}$ and the corresponding cosine amplitude, $p^{2\nu+1}$, which evolves as $p^{2\nu+1} + \alpha p^4$.[15] Unlike limiting the number of modes, we now exploit the fact that by increasing the momentum transfer $Q$, the dynamic structure becomes more sensitive to higher modes. In addition, for a given $Q$, the calculated *S(Q,t)/S(Q)* becomes independent of $p$, beyond a certain threshold ($p > p_{min}$ ). This uses the fact that $2\pi/Q$ probes a certain finite length, which limits the number of modes required to describe the experimental data theoretically. As a consequence, the spatial resolution is only determined by the $Q$ dependence of $S(Q,t)$ but not affected by the maximum $Q$.

The solid lines in Figures 4 (a) and (b) compare the result of our analysis with the experimental dynamical structure factor. We can accurately describe our experimental data by simultaneously fitting all the $Q$'s. From this analysis, we obtain the stiffness parameter α, that decreases with increasing molecular weight and/or temperature, cf. Table 1.

Based on this result, we can now estimate the minimum number of modes, $p_{min}$, that are required to theoretically describe the experimental $S(Q,t)$, within the $Q$ range of our NSE experiments, by solving $S(Q,t,\alpha = 0, N = p_{min}) = S(Q,t,\alpha,N = N_\infty)$. To calculate mode independent parameter α, above the threshold, $p > p_{min}$, we summed over $p$ = 1… 1000. We obtain considerably greater $p_{min}$ than earlier determined $p$ values. This $p_{min}$ is the maximum mode numbers which are visible in our experiment.

If we compare the quality of the fits based on the stiffness parameter (Figure 4) with those calculated assuming a low number of modes (Figure 3), we observe a similarly good description irrespective of their physical origins. The description of the relaxation of a chain by its mode spectrum assumes a certain number of statistically independent segments, connected by entropic springs. Numerous experiments justified the assumption of an infinite number of modes in case of flexible polymers like poly(ethylene-*alt*-propylene) or poly(ethylene glycol) (with $\alpha = 0$).[38,39] In the present case of conjugated polymer P3HT, the increased stiffness caused by the delocalized π-electron system introduces a finite correlation length, which decreases the number of statistically independent beads. Thus, the calculation of $S(Q,t)/S(Q)$ using a reduced number of modes is formally equivalent to the calculation using a stiffness parameter α (cf. Table 1). The absence of higher order modes elucidates the fact that the chain dynamics is partially frozen. Indeed, this is the first experimental evidence of the existence of single chain glass (SCG) state in a conjugated polymer. As the highest $Q$ is limited in experiments, $S(Q,t)$ cannot represent the entire mode spectrum. However, higher $Q$ values probe more local structures. If the stiffness already impacts the smaller momentum transfers, it is very likely that the wider angles would not change this discussion. However, we re-emphasize, if $Q$-values are reached that start to probe more local dynamics then deviations additional processes need to be incorporated in the model.[40,41] However, in the current situation there was no indication that this is the case with P3HT.

The comparison of the data with the Zimm model with all modes illustrates that the equivalent flexible polymer relaxes faster. At least two potential reasons can explain why the relaxation appears to be slower: (1) a reduced number of modes (Figure 3), or (2) damping of the modes (Figure 4). Apparently NSE data can be described by a finite number of modes (no damping). A decay of $S(Q,t)$ sets in, once modes contribute to the relaxation. Therefore, fewer modes result in less relaxation and more modes lead to a faster decay of $S(Q,t)$. However, the momentum transfer corresponds to a certain length-scale, $Q = 2\pi/d$. Therefore, the higher the $Q$ the more local the NSE experiment is, which implies higher modes. In a simplified wording, moving to the higher $Q$'s requires more modes contributing to $S(Q,t)$. In this context, we exploit the fact that each $Q$ has a maximum number of modes and increasing the number of modes would not change the calculated $S(Q,t)$ at this specific $Q^*$ and at every $Q < Q^*$. Obviously, this calculated $S(Q,t)$ relaxes faster than the experimental data. However, including damping slows down the decay. Therefore, we have now the opposite description.

This explanation can be rationalized by a simple estimation. For semi-flexible polymers, the number of modes, $p_{min}$ in equation 1, limits the displacement, $\cos(p_{min}\pi m/N)$, over $m = N/p_{min}$ segments. Therefore, we can estimate a dynamic rigid length, $R_{rigid}$. For distances less than $R_{rigid}$, the segments are correlated. These modes will be absent in the analysis. Thus, within a bead spring approach $R_{rigid}$ represents the length of a bead. It is given by: $R_{rigid} = \ell(N/p_{min})^{\nu} = R_{ee}N^{-\nu}(N/p_{min})^{\nu} = R_{ee}p_{min}^{-\nu}$.[15] From Table 1, it is evident that the effects of temperature and molecular weight are negligible on $R_{rigid}$, and we obtain $\langle R_{rigid}\rangle$ = 4.72±0.1 nm. From the structural standpoint, $R_{rigid}$ could likely be interpreted as the polymer conjugation length. Conjugation length is a length of a planarized chain segment where π-bonding is maintained over the entire segment, and is a key parameter which determines electronic and optoelectronic properties of conjugated polymers. Indeed, the value of $R_{rigid}$ corresponds to a bead length of approximately 12 thienyl repeating units, that is within the range of polythiophene conjugation length reported in literature (ranging between 10 and 20 repeating units).[42] It needs to be mentioned that the value of $R_{rigid}$ determined from the dynamic data is substantially higher than the P3HT persistence length (2.9 ± 0.1 nm) determined from wormlike chain modeling of static SANS data,[25] and reflects the fact that π-electron delocalization in P3HT extends on essentially longer distances than the geometrical persistence length.

It should be noted that, independently of the observed length scale, we obtained two significant parameters, namely, finite global stiffness, α and a finite size of the bead, $R_{rigid}$. The parameter α describes the damping of the mode relaxation. In the Rouse or Zimm approach, normal coordinates are introduced to solve the Langevin equation by simple exponential functions. The orthogonality of these normal coordinates follows from the uncorrelated random forces. This assumption corresponds to the freely jointed chain model that neglects correlations between bond vectors. In a good approximation, those finite correlations in a real polymer can be neglected if greater distances along the chain contour are considered. This leads to the introduction of $R_{rigid}$ and similarly to α.

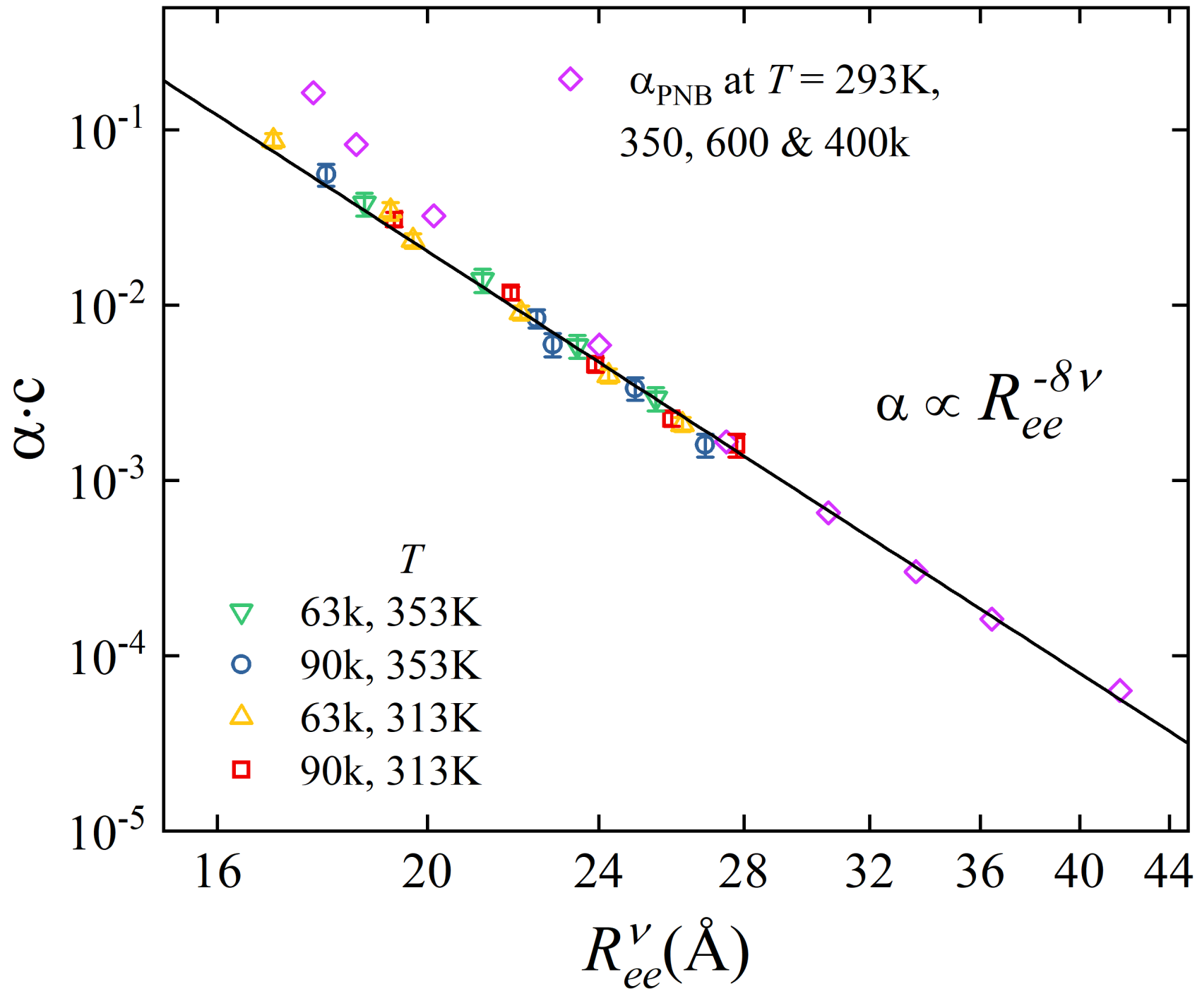

*Figure 5. Generalized scaling behavior, $\alpha \propto R_{ee}^{-8\nu}$, of the polymer stiffness, α, as a function of the chain dimension, $R_{ee}$ of P3HT samples (two molecular weights and two temperatures) and*

*polynorbornene (PNB) (three molecular weights from reference [15]). Here $\alpha$ is vertically scaled by a factor c.*

In order to investigate the scaling behavior between the chain end-to-end distance and the dynamical chain stiffness α, we systematically varied $R_{ee}$ from low to high values. As shown in Figure 5, we have used five different linearly spaced values above and below the experimentally obtained $R_{ee}$. This was done for both temperatures and polymer molecular weights. This reveals the dependence of α on the chain length. In addition to our results on P3HT, we have included the stiffness parameter $\alpha_{PNB}$ of polynorbornene (PNB) of different molecular weights in a good solvent.[15] For a better comparison, we rescaled $\alpha_{PNB}$ by a factor ~ 7. Irrespective of the polymer, molecular weight and temperature, we observe a generic power-law scaling, $\alpha \propto R_{ee}^{-8\nu}$. As a consequence, the molecular weight dependence of α is attributed to the increase in Gaussian coil dimension, $R_{ee}$ by a factor ~ 1.26.

We now want to explore how our findings based on the analysis of polymer dynamics, can be translated to macroscopic materials properties of conjugated polymers. As a special important case, we consider the correlation between the large-scale chain dynamics and thermochromism. Polythiophene shows a distinct thermochromic behavior both in solution and in solid state, as the polymer electronic absorption band undergoes reversible hypsochromic shift upon temperature increase.[43]

Let's sum up some of the essential facts. (i) The radius of gyration depends on the molecular weight as expected for a Gaussian coil, and increases around 15% with increase in temperature. At the same time, within the $Q$-range of our SANS experiments the aggregation is nearly independent of molecular weight or temperature. (ii) The bead size, $R_{rigid}$, is independent of molecular weight and temperature. (iii) The stiffness parameter, $\alpha$, decreases with increasing temperature and molecular weights. (iv) The absorption spectra of both P3HT samples in DCB-$D_4$ are independent of the molecular weight but show a thermochromic blue shift and an increase in band gap energy, $E_g$, with increasing temperature, cf. Figure 2 in the supplemental information (SI). These spectroscopic results agree with those found earlier for regioregular P3HT and seem to be common for semiconducting polymers.[44-48]

As it is widely accepted in the literature, the thermochromic blue shift in the absorption spectra of polythiophenes, including P3HT, upon increasing temperature is related to cooperative static conformational twisting (i.e. planar to non-planar conformational transition) of the π-electron conjugated backbone.[49-52] From our analysis, both the conjugation length (as reflected in the value of $R_{rigid}$) and our scaling law, $\alpha \propto R_{ee}^{-8\nu}$, show no dependence on temperature. It elucidates the fact that within the observed temperature range the constant bead size excludes a correlation with the observed changes in the absorption spectra. Also, the static chain end-to-end distance is not associated with the thermochromic blue-shift. Therefore, our results do not support static intramolecular conformational twisting of the π-conjugated backbone, and thus reduction of the conjugation length as a key factor in the observed thermochromic behavior.

The SANS data in Figure 2 cannot access the bead size since $\langle R_{rigid} \rangle$ = 4.7±0.1 nm corresponds to $Q = 2\pi/\langle R_{rigid} \rangle$ = 0.13 Å$^{-1}$, which is at the upper $Q$-limit of the SANS experiment. As the competition between coherent and incoherent scattering may contribute in this region, we abstain from the discussion of weak effects, which may not be related to the structure. Therefore, it impossible to see a structural peak. However, considering that our SANS data indicate significant aggregation of P3HT at low $Q$ (cf. SI) even at higher temperature, we should suggest that temperature-affected changes in the interchain aggregation may be responsible for the thermochromic blue-shift at the higher temperature. This finding emphasizes the unique role of the large-scale dynamics in understanding the fundamental physics of locally stiff polymers and deriving correlations between the chain stiffness and the macroscopic material properties, which has not been explored in the literature so far. We should emphasize that our findings derived from P3HT behavior in dilute solution are only applicable to the narrow temperature range (313 to 353 K) used in this study, and they may not be directly applied to thermochromism in solid state. Nevertheless, they do agree with recent conclusions about rather complicated nature of thermochromic phenomenon in conjugated polymers where multiple contributing factors are responsible for the observed spectroscopic changes.[53]

**Summary**

To conclude, we showed that conjugated polymers such as P3HT are ideally suited to understand the impact of locally stiff segments on the large-scale chain dynamics by SANS and NSE studies. This is the first experimental demonstration of single chain glassy state in conjugated polymers. We generalize the well-established Zimm model approach of flexible polymers and successfully describe the relaxation mode spectra of locally stiff chains. Only one parameter, the damping constant α, is sufficient to represent the full mode spectrum of both flexible and semiflexible chains. The increase in stiffness is reflected by a bead element $R_{rigid}$ of increased size, in concert with a reduction of the number of modes $p_{min}$. We derived a renormalized stiffness $\bar{\alpha}$ from the generic scaling of the stiffness $\alpha \propto R_{ee}^{-8\nu}$, and the molecular weight. Irrespective of the temperature, the band gap energy, and the molecular weight of the conjugated polymer, we obtain a universal behavior of the local chain stiffness. Our findings impressively confirm that the so-called local stiffness is the only controlling parameter to describe the dissipation of the entropic forces in large-scale polymer dynamics. As related to macroscopic materials properties, our results show a rather minor role of the conjugated backbone conformational twisting (planar to non-planar single-chain conformational transition leading to decrease in conjugation length) in the thermochromic behavior of P3HT, and indicate that interchain phenomena (such as change in interchain aggregation) and chain dynamics are likely responsible for the thermochromic phenomenon. We hypothesize that our findings may also be applicable for understanding of other related phenomena such as solvatochromic behavior of conjugated polymers where interplay of complex pathways has been recently shown to affect observed spectroscopic changes.[26] In this way, our findings open up new frontiers for understanding the macroscopic properties like viscoelastic and optoelectronic response for material processing as well as macromolecular crowding associated with the biological functioning of living organisms.

## Methods

**Sample preparation**. All reactions toward P3HT preparation were performed under an atmosphere of dry nitrogen, unless mentioned otherwise. Tetrahydrofuran (THF) for polymerization was dried by passing through activated alumina using a PS-400 Solvent Purification System from Innovative Technology, Inc. The water content of THF was periodically controlled by Karl Fischer titration, using a DL32 coulometric titrator from Mettler Toledo. Isopropylmagnesium chloride (2.0 M solution in THF) was purchased from Acros Organics. All other reagents and solvents were obtained from Sigma Aldrich and Alfa Aesar and used without further purification. Deuterated solvents (chloroform-D and 1,2-dichlorobenzene-$D_4$ (DCB-$D_4$)) were purchased from Cambridge Isotope Laboratories.

**Small angle neutron scattering (SANS) measurements.** SANS experiments were performed at the GP-SANS in High Flux Isotope Reactor (HFIR) at Oak Ridge National Laboratory (ORNL).[54] All the samples were measured in a standard 1 mm Hellma Banjo cells. The sample-to-detector distance (*d*) and the neutron wavelength ($\lambda$) were kept at $d$ = 19.2 m for $\lambda$ = 12 Å; $d$ = 8.8 m for $\lambda$ = 4.75 Å and $d$ = 1.1 m for $\lambda$ = 4.75 Å. This configuration covers a $Q$ - range from ~ 0.005 Å$^{-1}$ to ~ 0.23 Å$^{-1}$, where the momentum transfer, $Q = 4\pi sin(\theta/2)/\lambda$, for the scattering angle $\theta$. A wavelength resolution of $\Delta\lambda/\lambda$ = 15% was used. All data reduction into intensity $I(Q)$ vs. momentum transfer $Q = |\vec{Q}|$ was carried out following the standard procedures that are implemented in the SPICE SANS reduction package for the Igor software. The data scaling into absolute units (cm$^{-1}$), and the detector sensitivity correction was done with a porous silica standard measurement. The solvents and empty cell were measured separately as backgrounds and were subtracted.

**Neutron spin echo (NSE) measurements.** NSE spectroscopy was performed at the Spallation Neutron Source (SNS), ORNL, using the SNS-NSE spectrometer at BL-15.[55] We detect the normalized dynamic structure factor representing the sum of coherent ($S_{coh}$) and incoherent ($S_{inc}$) scattering. However, the coherent signal dominates, [56-58] i.e.,

$$\frac{S(Q,t)}{S(Q)} = \frac{\sigma_{coh}S_{coh}(Q,t) - \frac{1}{3}\sigma_{inc}S_{inc}(Q,t)}{\sigma_{coh}S_{coh}(Q) - \frac{1}{3}\sigma_{inc}S_{inc}(Q)} \approx \frac{S_{coh}(Q,t)}{S_{coh}(Q)} \tag{2}$$

Here, $\sigma_{coh}$ and, $\sigma_{inc}$ are the coherent and incoherent scattering intensities, respectively. For the NSE experiment an incoming wavelength band, $\Delta\lambda$, from 5 to 8 Å was used with 42 time channels for the time-of-flight data acquisition. This allowed to access a dynamic range of 2 ps ≤ $t$ ≤ 25 ns over a momentum transfer $Q$ = 0.062 – 0.124 $Å^{-1}$. For the measured coherent NSE data, corrections were performed using resolution data from $Al_2O_3$, sample and background from the DCB-$D_4$ solvent. The background subtraction was performed from the neutron spin-echo amplitude ($A$) to spin up-down intensity ratio ($Up - Dwn$) as described by Monkenbusch *et al.* [59]. We used specially designed two-part Al sample containers sealed with PTFE, attached to a tumbler, and maintaining a sample thickness of 4 mm. The data reduction was performed with the standard ECHODET software package of the SNS-NSE instrument. The incoherent and coherent contributions were determined by polarization analysis in the diffraction mode of the spectrometer. The elastic incoherent scattering from the background, including the solvent, the scattering that results from empty cell, sample environment and instrument, were subtracted accordingly to obtain the coherent dynamic structure factor. For further details the reader is referred to References.[56,59]

**NMR spectroscopy.** $^{1}$H NMR spectra were recorded at 400 MHz using Bruker AV-400 NMR spectrometer, and are reported in ppm downfield from tetramethylsilane.

**Acknowledgment**

The neutron scattering work was supported by the U.S. Department of Energy (DOE) under EPSCoR Grant No. DE-SC0012432 with additional support from the Louisiana Board of Regents. Research conducted at ORNL's High Flux Isotope Reactor (HFIR) and at Spallation Neutron Source (SNS) was sponsored by the Scientific User Facilities Division, Office of Basic Energy Sciences, U.S. Department of Energy (DoE). We thank Lilin He (HFIR), Marius Hofmann (LSU), Stefan Otto Huber (LSU) and Christopher Van Leeuwen (LSU) for helping with the scattering experiments.

**Disclaimer**

**Author contributions**

SG performed the NSE and SANS experiments. SC and EEN synthesized the samples and performed NMR and UV/vis experiments. PZ helped with the NSE experiments and data reduction. SG and GJS analyzed the NSE data. All authors contributed to discussion of the original idea of this work and writing the manuscript.

**Additional Information**

**Supplementary information**

Accompanies this paper. Figures are included to show temperature dependent viscosity of DCB, UV/vis absorbance spectra of P3HT as a function of temperature, and additional chain dynamics analysis based on NSE data.

**Competing Interests**: The authors declare no competing interests.

Email: g.sudipta26@gmail.com, een@niu.edu, gjschneider@lsu.edu